# Isotope shift of nuclear magnetic resonances in $^{129}$Xe and $^{131}$Xe caused by spin-exchange pumping by alkali metal atoms

V.I. Petrov[1], A.S. Pazgalev[2], and A.K. Vershovskii[2]

[1] Concern CSRI Elektropribor, JSC, 197046, St. Petersburg, Russia, 30 Malaya Posadskaya str.
[2] Ioffe Institute, 194021, St. Petersburg, Russia, 26 Politekhnicheskaya
e-mail: antver@mail.ioffe.ru

The effect of isotope shifts of nuclear magnetic resonance (NMR) frequency in xenon isotopes $^{129}$Xe and $^{131}$Xe polarized by optically-oriented alkali metal atoms is not only of fundamental interest, but also of practical significance since it is the main factor limiting the accuracy of a whole class of prospective navigation and metrological devices. We have studied the parametric dependences of the isotope shift and have shown that this effect is largely due to incomplete averaging of a inhomogeneous local magnetic field by two xenon isotopes. A numerical model has been derived, which qualitatively describes the effect of isotope frequency shift and provides a good quantitative agreement with the experiment.

## INTRODUCTION

Atomic nuclei for which direct optical pumping is not feasible can be polarized in the process of interaction with oriented alkali metal atoms (i.e., K, Rb, Cs, hereinafter referred to as Me)[1]. This process, known as spin-exchange pumping, shows particular efficiency for xenon due to the formation of weakly-coupled van der Waals molecules Me-Xe[2,3,4]. An additional two- or three-order amplification of the NMR signal in Xe is achieved due to the Fermi-contact hyperfine interaction with Me atoms[5].

Polarized xenon nuclei have found applications in medicine[6,7], navigation[8,9,10], precision measurement of the magnetic field[11], and search for fundamental interactions that do not preserve the local symmetry[12,13].

In[12], the magnetic resonance signal of odd xenon isotopes was used to search for long-range interactions involving axions and axion-like particles. The authors observed an effect which they called an isotope shift (IS). This effect shows itself in a mismatch of the values of the magnetic field induction measured by the $^{129}$Xe and $^{131}$Xe isotopes in a volume common to oriented rubidium atoms. Under these conditions, optically-oriented alkaline atoms create a local magnetic field in the cell volume. Contact interaction leads to an effective $k_{XeMe}$-fold increase in the magnetic field created by Xe nuclei and detected by Me atoms and to a $k_{MeXe} \approx k_{XeMe}$-fold increase of the field created by Me atoms and detected by Xe nuclei. In the subsequent discussion, we shall use the term "an *effective field*" to refer to the magnetic field strengthened by the contact interaction. The effective field created by oriented Me atoms in the cell will be called an internal field $B_a$.

In reference[5], there is no indication that the coefficients $k_{MeXe}$ may be different for the two xenon isotopes. Nevertheless, in the experiment[12], it was observed that the ratio of frequencies of NMR in xenon isotopes changed with the internal field $B_a$. The corresponding difference between the effective fields measured by the two isotopes was called an isotope shift. Quantitatively, the shift magnitude is described by the dimensionless variable $\delta B_A$ expressed through the ratio of the precession frequencies $f_i$ ($i$ = 1, 2 for $^{129}$Xe and $^{131}$Xe):

$$\frac{f_1}{f_2} = \rho\left(1 + \frac{\Delta B}{B_0}\right) = \rho\left(1 + \delta B_A \frac{P \cdot B_A}{B_0}\right), \quad (1)$$

where $\rho = |\gamma_1/\gamma_2|$, $\gamma_i$ ($i$ = 1,2) are the gyromagnetic ratios of isotopes; $P$ is the degree of polarization of the alkali metal atoms; $B_A$ is the effective field that the metal atoms would create in the case of complete ($P$ = 1) polarization; then the internal field is $B_a = P \cdot B_A$, and $\Delta B = \delta B_A \cdot B_a$.

The questions of the IS nature as well as of its dependence on the parameters of the experiment remain open. However, these questions are both of fundamental and applied significance since IS leads to the emergence of $\Delta B$ dependence on the internal field. Therefore, it has a decisive destructive effect on the accuracy of the whole class of prospective navigation devices using NMR in xenon[9,10,11,14,15].

Furthermore, being the most accurate instruments in their class, quantum metrological sensors are used to search for physical interactions that go beyond the Standard Model. Correct estimation of the systematic error of these devices determines the upper bound of the parameter domain of hypothetical interactions, and is the subject of a comprehensive analysis. Thus, reference[16] considered the factors determining the accuracy of the $^{129}$Xe-$^3$He-magnetometer. The precession frequencies of xenon and helium were compared in the absence of an effective field $B_a$. It was shown that the ratio of NMR frequencies is affected by the quadratic field gradient as well as the temperature gradient in the cell, which was explained by the fact that the temperature gradient causes a concentration gradient that turns out to be different for two atomic ensembles with mismatched diffusion constants. However, the estimates show that the IS effect recorded in[12] cannot be explained by these mechanisms since the difference between the diffusion constants of two xenon isotopes is obviously too small, just as the temperature gradient in the compact cell used in[12].

In[15], we demonstrated the dependence of the IS value on the concentration of the alkali metal and the value of the magnetic field gradient applied along the direction of optical pumping (z-axis). The following explanation for the IS effect was proposed:

- absorption of the pumping light along the z-direction by Me atoms leads to a nonuniform spatial distribution of pumping intensity $I_P(z)$ over the cell;
- polarization of Me atoms $P(z)$ is distributed according to the same law since the diffusion of Me can be neglected due to the short length of the diffusion path of oriented atoms as compared with the length of the cell;
- spin-exchange pumping of Xe isotopes takes place predominantly in the region of the highest alkaline metal polarization, that is, in the frontal area of the cell;
- the internal field $B_a(z)$ follows the distribution of P(z), which gives rise to the gradient $dB_a(z)/dz$.

Thus, in the region where xenon nuclei are oriented with maximum probability, they experience the effect of the local internal field that exceeds the average internal field. Diffusion of xenon atoms throughout the cell leads to an effective averaging of the spatially inhomogeneous field measured by atoms. The longer is the isotope relaxation time, the closer is the value of the field measured by it to the average across the cell.

The difference in the relaxation times of the two $T_i$ isotopes results in different averaging of the field, that is, emergence of an IS. In[15], we measured the IS dependence on temperature and showed that the IS effect vanishes at a temperature that provides an approximate equality of $T_i$. The effect of the internal gradient on xenon $T_i$ was previously described in[16] and it was later studied in[17]. In these publications, it was proposed to reduce the $^{129}$Xe relaxation, compensating for the internal field gradient using an external gradient; but the question of the effect produced by the field and its gradients on IS was not raised.

### EXPERIMENT

The experiment was carried out on a setup described in[15]. The setup was assembled in accordance with a two-beam scheme[18] with pumping by a circularly-polarized beam, resonant to the $D_1$ line of $^{133}$Cs and directed along the vector of the constant magnetic field $B_0$ (axis z). The EPR signal in cesium was detected by recording the rotation of the polarization plane of the transverse (i.e., directed along the x-axis) linearly-polarized beam detuned from the $D_1$ line of Cs. The cell under study contained Cs vapor, $N_2$, and natural xenon (which, in turn, contains 26.4% of $^{129}$Xe, 21.2% of $^{131}$Xe, and even zero-spin isotopes, acting as a buffer gas).

The magnetic field $B_0$ was stabilized at a level of 12 mkT; the field instability over the measurement time did not exceed a few unities of pT. The setup included a set of rings to create a linear magnetic field gradient along the pumping axis. Compared with[15], the set of rings was complemented with two compensating windings, which made it possible to eliminate errors caused by its effect on the sensor of the magnetic field stabilizer.

Optical pumping of cesium was alternately performed by the light of the left and right circular polarization; the pumping light intensity was ~ 10 mW. The precession of Xe nuclear moments was initiated with a π/2 pulse of a transverse resonant RF field. Switching of the pumping light polarization resulted in a change in the direction of the internal field vector, and thus, led to shifts in the isotope precession frequencies:

$$\begin{cases} f_1^+ = \frac{\gamma_1}{2\pi}\left[B_0 + \left(1 + \frac{\delta B_A}{2} + \frac{\kappa}{2}\right) \cdot B_a\right] \\ f_2^+ = \frac{\gamma_2}{2\pi}\left[B_0 + \left(1 - \frac{\delta B_A}{2} + \frac{\kappa}{2}\right) \cdot B_a\right] \\ f_1^- = \frac{\gamma_1}{2\pi}\left[B_0 - \left(1 + \frac{\delta B_A}{2} - \frac{\kappa}{2}\right) \cdot B_a\right] \\ f_2^- = \frac{\gamma_2}{2\pi}\left[B_0 - \left(1 - \frac{\delta B_A}{2} - \frac{\kappa}{2}\right) \cdot B_a\right] \end{cases} \quad (2)$$

where $f_i^-$ and $f_i^+$ are the precession frequencies measured at two pumping polarizations, and $\kappa \cdot B_a$ is a small ($\kappa \ll 1$) change in the effective field $B_a$ caused by the imperfection of the circular polarizer.

According to (1), based on the results of four frequency measurements, we can obtain $\delta B_A$, $B_a$, $\rho = \gamma_1/\gamma_2$ and $\kappa \cdot B_a$:

$$B_a = \frac{2B_0}{\frac{1}{\delta f_1 + \delta f_2} - \frac{\kappa}{4}} \approx 2B_0(\delta f_1 + \delta f_2) \quad (3)$$

$$\delta B_A = 2\frac{\delta f_1 - \delta f_2}{\delta f_1 + \delta f_2}, \quad (4)$$

where $\delta f_i = (f_i^+ - f_i^-)/(f_i^+ + f_i^-)$. From (2) and (3) it follows that $\kappa$ makes only a small contribution to the value of $B_a$ and does not make any contribution to the value of $\delta B_A$. The values of $\delta B_A$, $B_a$, $T_i$ were measured as a function of temperature in the range of 45–95°C (which corresponds approximately to a 50-fold change in the alkali density) and the gradient of the external magnetic field in the range of $dB_z/dz = -350...+350$ nT/cm.

The relaxation rates $G_i = 1/T_i$ obtained in the experiments were compared with the theory predictions[3,4] (Fig.1a) and with the results of our numerical simulation. As follows from Fig. 1b, IS nulling is observed at a temperature that provides an approximate equality of the isotope relaxation times: $T_0 = 79.5°C$. The inset in Fig. 1b shows that a rapid increase in $\delta B_A$ in the low-temperature region corresponds to a constant difference in the $\Delta B$ values of the fields measured by two isotopes: $\Delta B = (33 \pm 10)$ pT. This fact cannot be explained in the framework of our model; it will be discussed below.

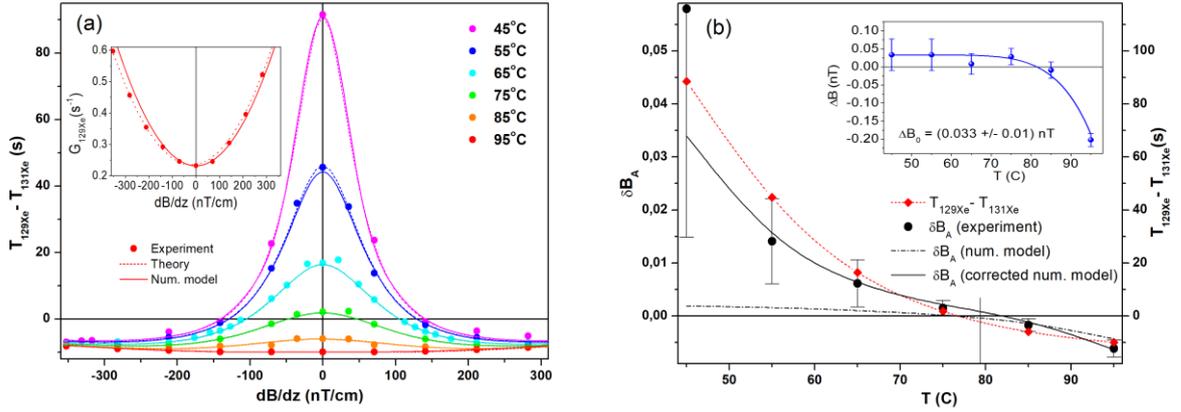

Fig. 1. a) Dependence of the difference between the relaxation times of isotopes $T_{129}$ and $T_{131}$ on the gradient of the external magnetic field at different temperatures. Inset: dependence of the $G_{129Xe}$ relaxation rate on the field gradient at 95°C. Solid lines: calculation of the $G_{129Xe}$ dependence on the field gradient according to[19]. Dotted line: the simulation results. b) Temperature dependence of $\delta B_A$ without the external field gradient. Circles: experiment. Dash-dotted line: numerical calculation. Solid line: the simulation results taking into account low-temperature factors. Short red dotted line: the difference between the relaxation times of $T_{129}$ and $T_{131}$. Inset: temperature dependence of the difference $\Delta B$ of the fields measured by Xe isotopes.

## NUMERICAL SIMULATION

We developed a simple one-dimensional model of the system, allowing for calculation of the temperature dependences of $\delta B_A$ and $G_i$ and the external gradient $dB_Z/dz$ (Fig. 2). The cell along the z-axis was divided by $N = 25..50$ $\Delta z = L/N$ thick layers. The distribution of all system parameters inside each layer was considered uniform. The pumping intensity $I_p(j)$ and the number of oriented cesium atoms $Ncs_j = Ncs(j)$ in each layer $j = 1..N$ were calculated taking into account the nonlinearity of absorption in the optically dense medium and the relaxation of Cs atoms on the cell walls. The simulation time step $\Delta t$ was taken equal to the diffusion time of Xe through a layer of thickness $\Delta z$: $\Delta t = \sqrt{\Delta z / D}$, where $D$ is the coefficient of Xe diffusion.

The distribution of polarized xenon nuclei $Nxe_{ij} = Nxe_i(j)$ was described by an array of complex numbers $|Nxe_{ij}| \cdot exp(i\varphi_{ij})$, where $\varphi_{ij}$ is the instantaneous precession phase of $Xe_i$ isotope in layer $j$. The phase evolution of one "generation" of a polarized nuclei ensemble was estimated numerically. It was assumed that the distribution of $Nxe_i(j)$ at the initial time $t = 0$ follows the distribution of $Ncs(j)$, and at subsequent points in time $(t = n \cdot \Delta t)$ evolves under the effect of diffusion, relaxation, external and internal magnetic fields, and their gradients. We took into consideration the fact that the relaxation of $^{131}$Xe also occurs on the cell surface as well as in the volume[1,2].

The evolution of the array $nXe_{i,j,n}$, at the calculation step $n$ is described by the three-diagonal evolution operator matrix. The structure of the operator reflects the fact that during the period of $\Delta t$, the diffusion process only binds the nearest adjacent layers. According to kinetic theory, the coefficients of isotope diffusion differ in accordance with the ratio of the roots of their masses; for xenon, $D_1/D_2 \approx 1.0077$. This difference was taken into account in the model, which, however, showed up only at the initial steps of the system and did not lead to any noticeable integral results.

At each step $n$, we calculated phase $\varphi_{i,n}$ of the $Xe_i$ atom precession averaged over the cell, taking into account the fact that the contribution of each layer to the common phase measured by Cs atoms is proportional to the number of these atoms in this layer.

The rate of change of the absolute value $|Nxe_{i,n}|$ was used to calculate $T_i$. Their dependence on the field gradient is consistent with the theory predictions with the accuracy of the coefficient close to unity[19]. The inset in Fig.1a shows that the dependence of the $^{129}$Xe relaxation rate has a minimum for a nonzero external gradient, which is due to the influence of the internal field linear gradient discovered earlier[14,16,17]. The numerical simulation (Fig. 1a) adequately describes the experimentally measured horizontal shift of this dependence.

The experiment also revealed a frequency shift that is not described by the proposed model. The discrepancy is especially noticeable at low temperatures $T \leq 65°C$, at which the value of the internal field is small. In this case, the difference in the $Xe_i$ precession frequencies is caused by a change in the sign of the circular polarization itself rather than by the internal field, and it was found to be $\Delta B = (33 \pm 10)$ pT. This effect may be explained by the asymmetry of the composite line of the $^{131}$Xe magnetic resonance.

In a cell that has an asymmetric shape, the $^{131}$Xe line is split into three components[20,21].

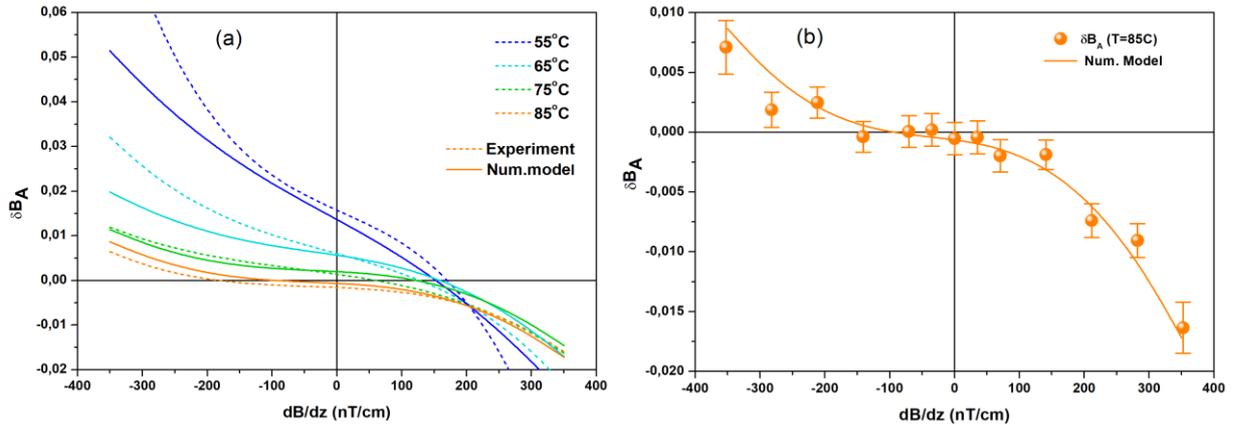

Fig. 2. Comparison of the numerical model with the experiment: (a) dotted line: experiment; solid line: numerical simulation; (b) δBA dependence on the external field gradient at T = 85°C. Solid line: numerical simulation.

When the amplitudes of the components are unequal, a change in the polarization sign can lead to a shift. The estimates show that a 2–3 percent difference in the cell size is sufficient to explain the effect; the same is true for the order of the asymmetry in the amplitudes of the components of the $^{131}$Xe quadrupole structure. The correction for the polarization-dependent shift of the $^{131}$Xe-magnetometer readings in the model eliminates the discrepancy with the experiment at low temperatures and has little or no effect on the results obtained at temperatures of 75°C and higher (Fig. 2b).

## CONCLUSIONS

We have studied the dependences of the isotopic shift (IS) and related parameters on temperature and the magnetic field gradient. A simple one-dimensional model has been developed, which describes the evolution of nuclear spins under the conditions of spin-exchange pumping with alkali metal atoms. It has been shown that in the temperature range of 65–85°C, which is of particular interest for compact NMR sensors, the model shows good agreement with the experiment, which confirms the validity of the explanation for the IS proposed earlier.

An anomalous IS was observed at low temperatures, and an explanation for it has been proposed. It has been shown that the isotopic shift can be eliminated to a large extent by equalizing the relaxation times of isotopes through an appropriate choice of the optimum temperature and/or compensation for the internal field gradient with the gradient of the external magnetic field.